\documentclass[11pt,a4paper]{article}

\usepackage[utf8]{inputenc}
\usepackage[T1]{fontenc}
\usepackage{lmodern}
\usepackage[margin=2.5cm]{geometry}
\usepackage{graphicx}
\usepackage{booktabs}
\usepackage{hyperref}
\usepackage[numbers,sort&compress]{natbib}
\usepackage{amsmath}
\usepackage{amssymb}
\usepackage{textcomp}
\usepackage{tikz}
\usepackage{xcolor}
\usepackage{subcaption}
\usepackage{multirow}
\usepackage{tabularx}
\usepackage{array}
\usepackage{enumitem}
\usepackage{microtype}

\usetikzlibrary{positioning,arrows.meta,shapes.geometric,fit,calc,backgrounds}

\definecolor{sourceblue}{HTML}{4472C4}
\definecolor{sourcegreen}{HTML}{548235}
\definecolor{parquetgray}{HTML}{7F7F7F}
\definecolor{duckdbyellow}{HTML}{FFC000}
\definecolor{xreforange}{HTML}{ED7D31}
\definecolor{ontopurple}{HTML}{7030A0}

\hypersetup{
  colorlinks=true,
  linkcolor=blue!70!black,
  citecolor=blue!70!black,
  urlcolor=blue!70!black,
}

\title{The Science Data Lake: A Unified Open Infrastructure\\ Integrating 293~Million Papers Across Eight Scholarly\\ Sources with Embedding-Based Ontology Alignment}

\author{
  Jonas Wilinski\thanks{Corresponding author.}\\
  \textit{Hamburg University of Technology (TUHH), Hamburg, Germany}\\
  \texttt{jonas.wilinski@tuhh.de}\\
  \small ORCID: \href{https://orcid.org/0009-0005-4672-7197}{0009-0005-4672-7197}
}

\date{March 2026}

\begin{document}

\maketitle
\begin{abstract}
Scholarly data are largely fragmented across siloed databases with divergent metadata and missing linkages among them. We present the Science Data Lake, a locally-deployable infrastructure built on DuckDB and simple Parquet files that unifies eight open sources---Semantic Scholar, OpenAlex, SciSciNet, Papers with Code, Retraction Watch, Reliance on Science, a preprint-to-published mapping, and Crossref---via DOI normalization while preserving source-level schemas. The resource comprises approximately 960\,GB of Parquet files spanning 293~million uniquely identifiable papers across 22~schemas and 153~SQL views. An embedding-based ontology alignment using BGE-large sentence embeddings maps 4,516 OpenAlex topics to 13~scientific ontologies (1.3~million terms), yielding 16,150 mappings covering 99.8\% of topics ($\geq 0.65$ threshold) with $F1 = 0.77$ at the recommended $\geq 0.85$ operating point, outperforming TF-IDF, BM25, and Jaro--Winkler baselines on a 300-pair gold-standard evaluation. We validate through 10~automated checks, cross-source citation agreement analysis (pairwise Pearson $r = 0.76$--$0.87$), and stratified manual annotation. Four vignettes demonstrate cross-source analyses infeasible with any single database. The resource is open source, deployable on a single drive or queryable
remotely via HuggingFace, and includes structured documentation suitable for large language model (LLM) based research agents.
\end{abstract}

\section{Background \& Summary}
\label{sec:background}

The advent of large-scale datasets tracing the workings of science has cultivated
a rapidly expanding ``science of science'' with its own data infrastructure,
metrics, and analytical frameworks~\cite{liu2023science}. Yet the databases that
underpin this field remain fragmented: Semantic Scholar provides influential
citation flags, open-access metadata, and full-text coverage for millions of
papers~\cite{kinney2023s2ag}; OpenAlex---the most comprehensive open index to
date---offers field-weighted citation impact (FWCI), a four-level topic taxonomy
with 4,516~leaf topics, geocoded institution affiliations for 121\,K
institutions, and 11.7~million funding awards with dollar amounts, all under a
CC0 license~\cite{priem2022openalex}; SciSciNet contributes a landmark
collection of pre-computed science-of-science metrics including disruption
indices, atypicality scores, and patent linkages~\cite{lin2023sciscinet}; Papers
with Code links papers to reproducible code, though the platform has ceased
active operations, making its archived snapshot a non-renewable
resource~\cite{paperswithcode2024}; Retraction Watch tracks integrity
events~\cite{retractionwatch2024}; and Reliance on Science maps patent-to-paper
citations~\cite{marx2020ros}. Each of these sources represents a substantial
achievement in opening the scientific record; however, no single source captures
all facets simultaneously. Researchers who wish to combine them---for example, to
study whether disruptive papers are more likely to release code, or whether
retracted papers show anomalous citation patterns across databases---must write
ad-hoc integration scripts that are rarely shared or reproduced.

A systematic evaluation of 59 scholarly databases found substantial variation in
backward and forward citation coverage~\cite{gusenbauer2024citation}. Large-scale
pairwise comparisons of bibliographic data sources have revealed non-trivial
differences in metadata, document types, and citation
counts~\cite{visser2021comparison}, but record-level joins across more than two
sources remain uncommon. The lack of a shared infrastructure forces each research
group to repeat the same data-wrangling steps, wasting effort and introducing
inconsistencies.

Several systems have begun to address this gap (Table~\ref{tab:comparison}).
SciSciNet~\cite{lin2023sciscinet} provides a rich ``data lake'' built on
Microsoft Academic Graph (now OpenAlex) with pre-computed science-of-science
metrics and linkages to patents, grants, and clinical trials, but its
bibliometric backbone draws from a single index and it does not preserve
independent source-level schemas for cross-source comparison.
PubGraph~\cite{ahrabian2023pubgraph} merges Wikidata, OpenAlex, and Semantic
Scholar into a unified knowledge graph using the Wikidata ontology, but
collapses source-level schemas, sacrificing the ability to compare how different
sources describe the same paper. SemOpenAlex~\cite{farber2023semopenalex}
re-encodes OpenAlex as 26~billion Resource Description Framework (RDF)
triples, offering semantic-web
interoperability but remaining a single-source resource.
Dimensions~\cite{hook2018dimensions} provides SQL-queryable access to a
comprehensive commercial database, but its proprietary nature limits
reproducibility. The Open Research Knowledge Graph
(ORKG)~\cite{jaradeh2019orkg} takes a complementary approach, focusing on
structured descriptions of \emph{research contributions} (methods, results,
datasets) as curated RDF triples rather than bibliometric metadata; it
federates metadata from Crossref, Semantic Scholar, and OpenAIRE but does
not provide record-level citation counts, scientometric indicators, or
cross-source comparison of the same paper across independent databases.

\begin{table}[t]
\centering
\caption{Comparison with existing scholarly data integration systems.}
\label{tab:comparison}
\footnotesize
\setlength{\tabcolsep}{3pt}
\begin{tabularx}{\textwidth}{@{} l c c c c X @{}}
\toprule
\textbf{System} & \textbf{Src} & \textbf{Multi} & \textbf{Schemas} & \textbf{Open} & \textbf{Key limitation} \\
\midrule
SciSciNet~\cite{lin2023sciscinet}        & 1+ & \texttimes & ---      & \checkmark & Single bibliometric backbone (MAG/OpenAlex); no independent cross-source comparison \\[2pt]
PubGraph~\cite{ahrabian2023pubgraph}     & 3  & \checkmark & Merged   & \checkmark & Collapses into unified Wikidata schema; source-level provenance lost \\[2pt]
SemOpenAlex~\cite{farber2023semopenalex} & 1  & \texttimes & ---      & \checkmark & Single source (OpenAlex only); requires SPARQL expertise \\[2pt]
Dimensions~\cite{hook2018dimensions}     & 1  & \texttimes & ---      & \texttildelow & Commercial; limited programmatic access; not fully reproducible \\[2pt]
ORKG~\cite{jaradeh2019orkg}              & 3+ & \checkmark & KG       & \checkmark & Focuses on curated research contributions, not bibliometric record-level comparison \\[2pt]
\textbf{Science Data Lake}               & \textbf{8} & \checkmark & \textbf{Preserved} & \checkmark & Requires {\raise.17ex\hbox{$\scriptstyle\sim$}}1\,TB local storage; snapshot-based \\
\bottomrule
\end{tabularx}
\end{table}

Rather than attempting to replace these individually excellent resources, the
Science Data Lake provides an open infrastructure foundation for
science-of-science research---analogous to what
SciSciNet~\cite{lin2023sciscinet} achieved for single-source metrics, but
extended across eight independent data sources. The resource makes three
contributions. \textbf{First}, a \emph{multi-source preserving architecture}
that integrates eight open scholarly databases into a single resource while
retaining each source's native schema, enabling direct cross-source comparison at
the record level. Because upstream sources---particularly OpenAlex---evolve their
schema across snapshot partitions, the conversion pipeline employs dynamic Python
scripts that auto-discover entity types and column structures rather than relying
on hard-coded schemas, ensuring the infrastructure remains current as sources
change. \textbf{Second}, an \emph{embedding-based ontology alignment}
method that bridges OpenAlex's flat topic taxonomy to 13~formal scientific
ontologies using BGE-large sentence embeddings~\cite{xiao2024bge}, achieving
$F1 = 0.77$ at the recommended operating point and outperforming TF-IDF,
BM25, and Jaro--Winkler baselines on a gold-standard evaluation. \textbf{Third}, a
\emph{cross-source record-level comparison layer} (\texttt{unified\_papers},
293M rows, 29~columns) that enables simultaneous queries across all
sources---supporting analyses such as multi-database citation reliability
assessment that no single source or pairwise comparison can provide.
Beyond traditional research workflows, the SQL-native architecture and
structured documentation make the data lake particularly amenable to
emerging LLM-based research agents~\cite{wang2023aiforscience}: a
machine-readable schema reference is designed to facilitate navigation of
the 153-view, 22-schema structure.

\section{Methods}
\label{sec:methods}

\subsection{Data Sources}
\label{sec:sources}

The Science Data Lake integrates eight open scholarly data sources, each contributing
distinct metadata facets (Table~\ref{tab:sources}).

\begin{table}[t]
\centering
\caption{Overview of integrated data sources. Record counts reflect the snapshot
versions used in this release.}
\label{tab:sources}
\small
\begin{tabularx}{\textwidth}{@{}l r l X@{}}
\toprule
\textbf{Source} & \textbf{Records} & \textbf{License} & \textbf{Key content} \\
\midrule
Semantic Scholar (S2AG)     & 231M  & ODC-BY   & Citations, influential citations, open access \\
OpenAlex                    & 479M  & CC0      & FWCI, topics, types, languages, funding awards \\
SciSciNet                   & 250M  & CC BY    & Disruption, atypicality, team size \\
Papers with Code            & 513K  & CC BY-SA & Code repositories, tasks, datasets \\
Retraction Watch            & 69K   & Open     & Retraction reasons and dates \\
Reliance on Science (RoS)   & 47.8M & CC BY-NC & Patent--paper citation pairs \\
Preprint-to-Published (P2P) & 146K  & Open     & bioRxiv/medRxiv DOI to published DOI \\
Crossref                    & ---\textsuperscript{a}   & Open     & DOI metadata, reference lists \\
\bottomrule
\end{tabularx}
\vspace{2pt}
{\footnotesize \textsuperscript{a}Crossref serves as a metadata enrichment and DOI validation layer rather than a standalone record source; no independent record count is attributed.}
\end{table}

\textbf{Semantic Scholar Academic Graph (S2AG)}~\cite{kinney2023s2ag} provides
bibliometric metadata for approximately 231~million papers, including citation counts,
influential citation counts (based on citation context analysis), and open-access
status flags.

\textbf{OpenAlex}~\cite{priem2022openalex} is an open catalogue of 479~million
scholarly works with field-weighted citation impact (FWCI), a four-level topic taxonomy
(domain, field, subfield, topic with 4,516~leaf topics), document types, and language
annotations.

\textbf{SciSciNet}~\cite{lin2023sciscinet} augments OpenAlex records with
pre-computed science-of-science metrics including the disruption index
CD\textsubscript{5}~\cite{funk2017disruption,wu2019disruption}, journal atypicality
$z$-scores~\cite{uzzi2013atypicality}, team size indicators, and patent citation
counts for 250~million papers.

\textbf{Papers with Code}~\cite{paperswithcode2024} links 513~thousand machine-learning
papers to their associated code repositories, benchmark tasks, and datasets.

\textbf{Retraction Watch}~\cite{retractionwatch2024} catalogues approximately 69~thousand
retracted or corrected publications with structured retraction reasons and dates.

\textbf{Reliance on Science (RoS)}~\cite{marx2020ros} provides 47.8~million patent-to-paper citation pairs from global patent
offices, with confidence scores and citation location metadata.

\textbf{Preprint-to-Published (P2P)} provides approximately 146~thousand mappings from
bioRxiv and medRxiv preprint DOIs to their corresponding published-version DOIs.

\textbf{Crossref} contributes DOI metadata and reference lists used for DOI
validation and supplementary linkage.

\begin{figure}[t]
\centering
\includegraphics[width=\textwidth]{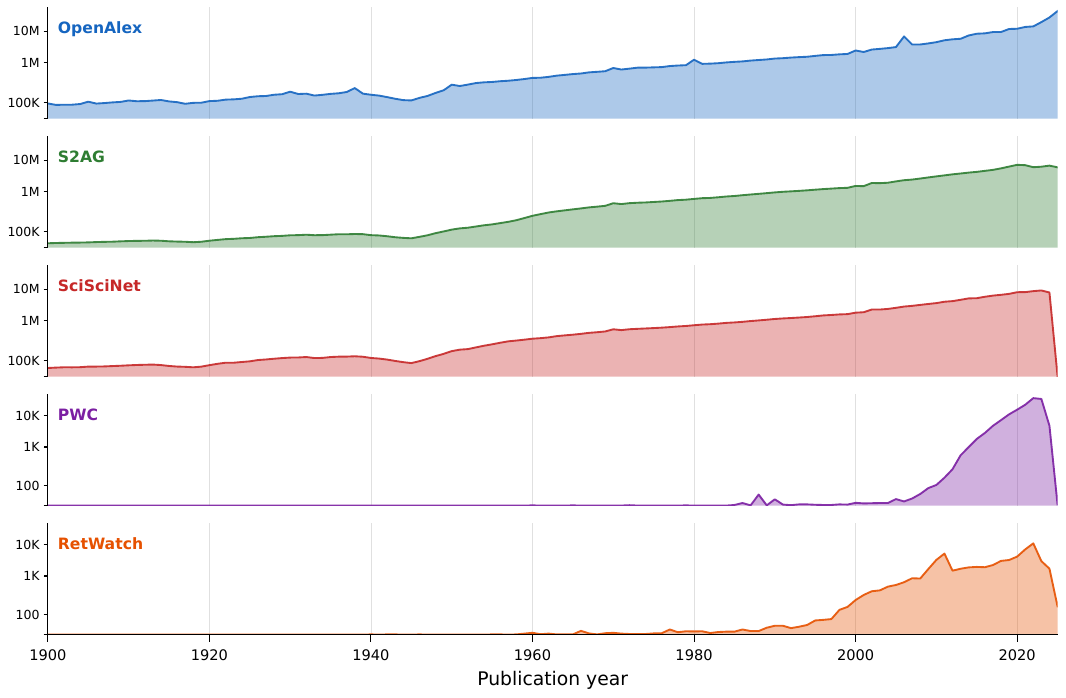}
\caption{Temporal coverage by source (symlog scale). Publication-year distributions
for DOI-linked records in the unified index (see Table~\ref{tab:sources} for full
source sizes). The symlog y-axis is linear near zero and logarithmic above, revealing
both historical depth and recent growth. OpenAlex, S2AG, and SciSciNet extend back to
1900 with tens of thousands of papers per year, while all three show clear acceleration
after $\sim$1960. SciSciNet exhibits a sharp cutoff after $\sim$2022 when its metrics
computation ends. The specialized sources (PWC, Retraction Watch) cover narrow temporal
windows concentrated in the last two decades.}
\label{fig:temporal}
\end{figure}

Figure~\ref{fig:temporal} shows the temporal coverage of each source, revealing
structural differences: OpenAlex extends back several centuries, S2AG is concentrated
in recent decades with a computer-science emphasis, and SciSciNet metrics end around
2022.

\subsection{Architecture}
\label{sec:architecture}

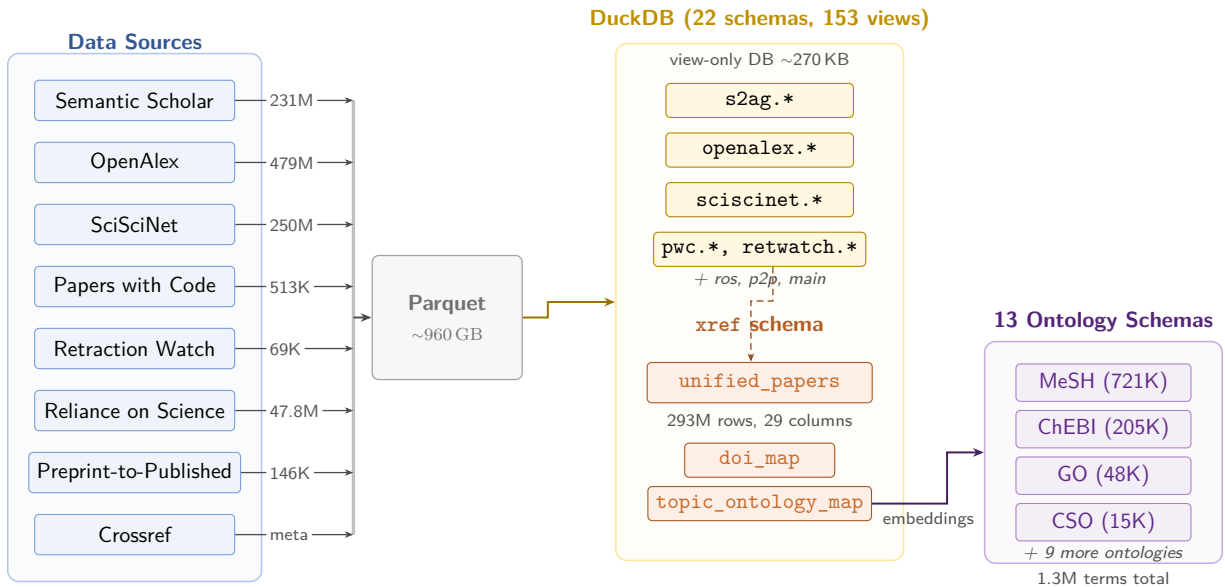
\begin{figure}[t]
\centering
\resizebox{\textwidth}{!}{%
\begin{tikzpicture}[
  source/.style={
    rectangle, rounded corners=2pt, minimum width=3.2cm, minimum height=0.65cm,
    draw=sourceblue!70, fill=sourceblue!8, font=\small\sffamily,
    text=black, line width=0.5pt, align=center
  },
  countlabel/.style={font=\scriptsize\sffamily\color{gray!70!black}, anchor=west,
    fill=white, inner sep=1.5pt, rounded corners=1pt},
  parquetnode/.style={
    rectangle, rounded corners=3pt, minimum width=2.4cm, minimum height=2.0cm,
    draw=parquetgray!70, fill=parquetgray!6, font=\small\sffamily\bfseries,
    text=parquetgray!90!black, line width=0.6pt, align=center
  },
  schemanode/.style={
    rectangle, rounded corners=2pt, minimum width=3.0cm, minimum height=0.55cm,
    draw=duckdbyellow!70!black, fill=duckdbyellow!12,
    font=\small\ttfamily, text=black, line width=0.5pt
  },
  xrefnode/.style={
    rectangle, rounded corners=2pt, minimum width=2.8cm, minimum height=0.55cm,
    draw=xreforange!70!black, fill=xreforange!12,
    font=\small\ttfamily, text=xreforange!85!black, line width=0.5pt
  },
  ontonode/.style={
    rectangle, rounded corners=2pt, minimum width=2.8cm, minimum height=0.55cm,
    draw=ontopurple!60, fill=ontopurple!8, font=\small\sffamily,
    text=ontopurple!85!black, line width=0.5pt
  },
  groupframe/.style={
    rectangle, rounded corners=5pt, draw=#1!40, fill=#1!3,
    line width=0.7pt, inner sep=8pt
  },
  grouplabel/.style={font=\small\sffamily\bfseries, text=#1!75!black},
  myarrow/.style={-{Stealth[length=4pt,width=3pt]}, line width=0.6pt, color=#1!60!black},
  thickarrow/.style={-{Stealth[length=5pt,width=4pt]}, line width=0.9pt, color=#1!60!black},
  note/.style={font=\scriptsize\sffamily, text=gray!65!black},
]

\node[source] (s2ag)    at (0, 0)    {Semantic Scholar};
\node[source] (oalex)   at (0,-1.0)  {OpenAlex};
\node[source] (ssn)     at (0,-2.0)  {SciSciNet};
\node[source] (pwc)     at (0,-3.0)  {Papers with Code};
\node[source] (retw)    at (0,-4.0)  {Retraction Watch};
\node[source] (ros)     at (0,-5.0)  {Reliance on Science};
\node[source] (p2p)     at (0,-6.0)  {Preprint-to-Published};
\node[source] (xrefsrc) at (0,-7.0)  {Crossref};

\node[grouplabel=sourceblue, anchor=south] at (0, 0.7) {Data Sources};


\coordinate (bustop)    at (3.5, 0);
\coordinate (busbottom) at (3.5, -7.0);
\draw[parquetgray!50, line width=1.5pt] (bustop) -- (busbottom);

\foreach \src in {s2ag,oalex,ssn,pwc,retw,ros,p2p,xrefsrc} {
  \draw[myarrow=parquetgray] (\src.east) -- (3.5, 0 |- \src.east);
}

\node[countlabel] at (2.1, 0.0)    {231M};
\node[countlabel] at (2.1,-1.0)  {479M};
\node[countlabel] at (2.1,-2.0)  {250M};
\node[countlabel] at (2.1,-3.0)  {513K};
\node[countlabel] at (2.1,-4.0)  {69K};
\node[countlabel] at (2.1,-5.0)  {47.8M};
\node[countlabel] at (2.1,-6.0)  {146K};
\node[countlabel] at (2.1,-7.0)  {meta};

\node[parquetnode] (pq) at (5, -3.5)
  {Parquet\\[2pt]\normalfont\scriptsize{\raise.17ex\hbox{$\scriptstyle\sim$}}960\,GB};

\draw[thickarrow=parquetgray] (3.5, -3.5) -- (pq.west);


\node[grouplabel=duckdbyellow, anchor=south] at (10.0, 1.0)
  {DuckDB (22 schemas, 153 views)};
\node[note, anchor=north] at (10.0, 0.9)
  {view-only DB {\raise.17ex\hbox{$\scriptstyle\sim$}}270\,KB};

\node[schemanode] (ds1) at (10.0, 0.0)  {s2ag.*};
\node[schemanode] (ds2) at (10.0,-0.80) {openalex.*};
\node[schemanode] (ds3) at (10.0,-1.60) {sciscinet.*};
\node[schemanode] (ds4) at (10.0,-2.40) {pwc.*, retwatch.*};
\node[note]       at (10.0,-2.90)       {\textit{+ ros, p2p, main}};


\node[grouplabel=xreforange, anchor=south] at (10.0, -3.85)
  {\texttt{xref} schema};

\node[xrefnode, minimum width=3.6cm, minimum height=0.65cm] (unified) at (10.0, -4.55)
  {unified\_papers};
\node[note, below=1pt of unified] {293M rows, 29 columns};

\node[xrefnode, minimum width=2.4cm] (doimap) at (10, -5.8) {doi\_map};
\node[xrefnode, minimum width=3.0cm] (tomap)  at (10, -6.5) {topic\_ontology\_map};


\node[grouplabel=ontopurple, anchor=south] at (15.5, -3.85)
  {13 Ontology Schemas};

\node[ontonode] (mesh)  at (15.5, -4.55) {MeSH (721K)};
\node[ontonode] (chebi) at (15.5, -5.30) {ChEBI (205K)};
\node[ontonode] (go)    at (15.5, -6.05) {GO (48K)};
\node[ontonode] (cso)   at (15.5, -6.80) {CSO (15K)};
\node[note]     at (15.5, -7.30) {\textit{+ 9 more ontologies}};
\node[note]     at (15.5, -7.70) {1.3M terms total};

\begin{scope}[on background layer]
  \node[groupframe=sourceblue,
    fit=(s2ag)(xrefsrc), inner sep=12pt] {};

  \node[groupframe=duckdbyellow,
    fit=(ds1)(ds4)(unified)(doimap)(tomap),
    inner sep=14pt, inner ysep=18pt] (duckdb) {};

  \node[groupframe=ontopurple,
    fit=(mesh)(cso), inner xsep=14pt, inner ysep=10pt] (ontology) {};
\end{scope}


\draw[thickarrow=duckdbyellow] (pq.east) -- ++(0.4,0)
  node[above, note, pos=0.5] {} |- (duckdb.west);


\draw[myarrow=xreforange, densely dashed]
  ([xshift=6pt]ds4.south) -- ++(0,-0.55) -| ([xshift=-4pt]unified.north);

\draw[thickarrow=ontopurple] (tomap.east) -- ++(1.2,0) |- (ontology.west);
\node[note, anchor=north] at (12.7, -6.5) {embeddings};

\end{tikzpicture}%
}
\caption{Architecture of the Science Data Lake.
Eight open scholarly data sources (left) are converted to Apache Parquet format
({\raise.17ex\hbox{$\scriptstyle\sim$}}960\,GB) and exposed as SQL views through
a lightweight DuckDB database (center). Each source retains its native schema for
source-level fidelity. The cross-referencing \texttt{xref} schema (orange) links
records via DOI normalization (\texttt{unified\_papers}, 293M rows) and connects
OpenAlex topics to 13~scientific ontologies (right) through BGE-large
embedding-based alignment.}
\label{fig:architecture}
\end{figure}

The Science Data Lake is built on a \emph{views-over-Parquet} architecture using
DuckDB~\cite{raasveldt2019duckdb} (Figure~\ref{fig:architecture}). Each data source
is first converted from its native format (JSON Lines, CSV, N-Triples) into columnar
Apache Parquet files, totaling approximately 960\,GB on disk. A lightweight DuckDB
database ({\raise.17ex\hbox{$\scriptstyle\sim$}}270\,KB) defines 153~SQL views
organized into 22~schemas that reference these Parquet files without copying data.

The schema design follows two principles. \textbf{Source-level preservation}: each
data source retains its native schema within a dedicated namespace (e.g.,
\texttt{s2ag.papers}, \texttt{openalex.works}, \texttt{sciscinet.paper\_metrics}),
enabling direct inspection of how different databases represent the same paper.
\textbf{Cross-referencing via the \texttt{xref} schema}: three materialized views
bridge across sources---\texttt{doi\_map} (DOI normalization), \texttt{unified\_papers}
(293M-row join table), and \texttt{topic\_ontology\_map} (ontology alignment).

We chose DuckDB over traditional server-based systems such as PostgreSQL for
three reasons: its embeddable, serverless design requires no installation or
administration; its columnar engine is optimized for the online analytical processing (OLAP) query
patterns typical of bibliometric research; and it reads Parquet
files directly through SQL views without importing data, keeping the database
file under 300\,KB while exposing nearly 1\,TB of data.

The system supports dual-mode access: local deployment on a high-capacity drive
for full-speed analytical queries, or remote access through HuggingFace-hosted
Parquet files for users without local storage.

A reproducible pipeline orchestrated by a master CLI script (\texttt{datalake\_cli.py})
automates the full workflow: downloading source snapshots, converting to Parquet,
creating DuckDB views, materializing cross-reference tables, and building the ontology
linkage.

\subsection{DOI Normalization and Record Linkage}
\label{sec:doi}

Digital Object Identifiers (DOIs) serve as the primary key for cross-source record
linkage, but sources store them in incompatible formats (Table~\ref{tab:doi}).

\begin{table}[t]
\centering
\caption{DOI format differences across data sources and the normalization applied.}
\label{tab:doi}
\small
\begin{tabular}{@{}l l l@{}}
\toprule
\textbf{Source} & \textbf{Raw DOI format} & \textbf{Normalization} \\
\midrule
S2AG             & lowercase, no prefix (\texttt{10.1038/\ldots}) & Canonical (none) \\
OpenAlex         & lowercase, \texttt{https://doi.org/} prefix    & Strip prefix \\
SciSciNet        & lowercase, \texttt{https://doi.org/} prefix    & Strip prefix \\
Papers with Code & lowercase, no prefix                           & None \\
Retraction Watch & lowercase, no prefix                           & None \\
Crossref         & mixed case                                     & Lowercase \\
\bottomrule
\end{tabular}
\end{table}

All DOIs are normalized to a canonical lowercase, prefix-free format. The
\texttt{xref.doi\_map} view implements this normalization as a union of
source-specific sub-queries, each applying the appropriate transformation.

The resulting \texttt{xref.unified\_papers} table contains 293,123,121 unique DOIs
with 29~columns drawn from all sources, including six Boolean coverage flags
indicating which sources contain each paper. Table~\ref{tab:overlap} summarizes the
pairwise coverage.

\begin{table}[t]
\centering
\caption{Cross-source coverage of the 293M unified papers. Each cell shows the
percentage of papers present in the column source that are also present in the
row source.}
\label{tab:overlap}
\small
\begin{tabular}{@{}l r r r r r r@{}}
\toprule
 & \textbf{OpenAlex} & \textbf{SciSciNet} & \textbf{S2AG} & \textbf{PWC} & \textbf{RetWatch} & \textbf{RoS} \\
\midrule
Coverage (\%) & 99.67 & 54.08 & 45.52 & 0.048 & 0.020 & 0.19 \\
\bottomrule
\end{tabular}
\end{table}

OpenAlex provides the broadest coverage at 99.67\% of all DOIs, consistent with
its role as a comprehensive open index. SciSciNet and S2AG cover approximately
half the DOI space, reflecting their focus on papers with sufficient citation
data for metric computation. The specialized sources (Papers with Code, Retraction
Watch, Reliance on Science) contribute smaller but unique record sets that cannot
be obtained from the three large databases.

\begin{figure}[t]
\centering
\includegraphics[width=\textwidth]{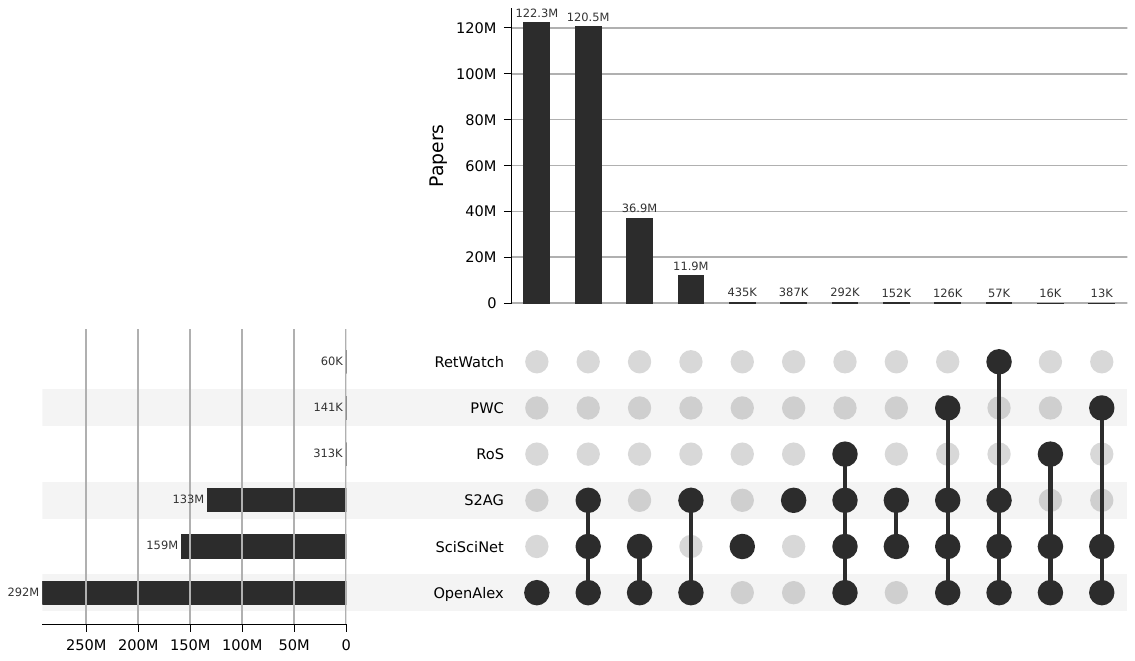}
\caption{UpSet plot showing the intersection structure across six data sources.
Bars represent the number of papers in each source combination. Of 34~observed
combinations, the three-way overlap of OpenAlex, SciSciNet, and S2AG accounts
for the largest multi-source intersection.}
\label{fig:upset}
\end{figure}

Figure~\ref{fig:upset} shows the UpSet plot of the six-source overlap, revealing
34~observed source combinations. The dominant combination is OpenAlex-only
(45.0\%), followed by the three-way overlap of OpenAlex, SciSciNet, and S2AG
(38.2\%).

\subsection{Embedding-Based Ontology Alignment}
\label{sec:ontology}

OpenAlex assigns papers to a flat topic taxonomy of 4,516~topics organized into
four hierarchical levels (252~subfields, 26~fields, 4~domains), but these topics
lack mappings to formal scientific ontologies that encode domain-specific
knowledge. To bridge this gap, we developed an embedding-based alignment method
that maps OpenAlex topics to 13~scientific ontologies comprising 1.3~million
terms in total.

The 13~ontologies span diverse scientific domains: Medical Subject Headings
(MeSH; 721K terms)~\cite{lipscomb2000mesh}, Chemical Entities of Biological
Interest (ChEBI; 205K)~\cite{hastings2016chebi}, NCI Thesaurus (NCIT;
204K)~\cite{sioutos2007ncit}, Gene Ontology (GO;
48K)~\cite{go2021geneontology}, AGROVOC
(42K)~\cite{caracciolo2013agrovoc}, Computer Science Ontology (CSO;
15K)~\cite{salatino2020cso}, Disease Ontology
(DOID)~\cite{schriml2022doid}, Human Phenotype Ontology
(HPO)~\cite{kohler2021hpo}, EDAM bioinformatics
ontology~\cite{ison2013edam}, UNESCO
Thesaurus\footnote{\url{https://vocabularies.unesco.org/browser/thesaurus/en/}},
Standard Thesaurus for Economics
(STW)\footnote{\url{https://zbw.eu/stw/}}, Physics Subject Headings
(PhySH)\footnote{\url{https://physh.org/}}, and the Mathematics Subject
Classification (MSC2020)\footnote{\url{https://msc2020.org/}}. Each ontology was converted
from its native format (Open Biomedical Ontologies (OBO), Simple Knowledge
Organization System (SKOS)/Resource Description Framework (RDF), N-Triples,
CSV) to a uniform Parquet representation using format-specific parsers, and
simultaneously loaded into an Oxigraph RDF triple store for SPARQL queries.

We employed a hybrid alignment strategy. For the 10~smaller ontologies (291K
terms including synonyms), we computed dense embeddings using the BAAI General Embedding
(BGE) large-en-v1.5 model~\cite{xiao2024bge} (335M parameters, 1024
dimensions) and performed nearest-neighbour search via a Facebook AI
Similarity Search (FAISS) index~\cite{johnson2021faiss}. For the three largest ontologies (MeSH, ChEBI, NCIT), which
together account for 1.1M terms and would dominate the embedding space, we used
exact string matching to ensure precision.

Table~\ref{tab:ontology_tiers} summarizes the alignment quality with
representative examples for each tier. At the exact-match tier ($\geq 0.95$),
topics align with near-identical ontology terms---for instance, the OpenAlex
topic ``Machine Learning'' maps to CSO's ``machine learning'' (similarity
0.98). At the high-quality tier ($\geq 0.85$), semantically related but
differently named concepts are captured---e.g., ``Artificial Intelligence in
Medicine'' maps to EDAM's ``Medical informatics'' (0.87). Relaxing to
$\geq 0.65$ yields 16,150~mappings covering 4,509 of 4,516 topics (99.84\%),
including broader associations such as ``Soil Chemistry'' mapping to
AGROVOC's ``soil chemicophysical properties'' (0.68).

\begin{table}[t]
\centering
\caption{Ontology alignment quality tiers. Each tier includes all mappings at or
above the similarity threshold.}
\label{tab:ontology_tiers}
\small
\begin{tabular}{@{}l r r r@{}}
\toprule
\textbf{Quality tier} & \textbf{Similarity} & \textbf{Mappings} & \textbf{Topics covered} \\
\midrule
Exact match   & $\geq 0.95$ &     85 & 71 (1.6\%) \\
High quality  & $\geq 0.85$ &  2,527 & 1,647 (36.5\%) \\
All           & $\geq 0.65$ & 16,150 & 4,509 (99.84\%) \\
\bottomrule
\end{tabular}
\end{table}

To contextualize the embedding approach, we compared it against three
standard baselines (Table~\ref{tab:baselines}): character-level TF-IDF
cosine similarity (bigram-to-4-gram, \texttt{char\_wb} analyzer),
BM25~\cite{robertson2009bm25} (Okapi BM25 over whitespace-tokenized labels,
min-max normalized, top-20 per topic), and Jaro--Winkler string similarity
at a threshold of 0.90. All methods operate on the same 4,516 topics and
291K ontology terms, with large ontologies (MeSH, ChEBI, NCIT) using exact
matching in all cases for consistency.

\begin{table}[t]
\centering
\caption{Comparison of ontology alignment methods. All methods use the same
topic and ontology term sets. Mappings and topic counts are at each method's
operating threshold. Precision (P), recall (R), and F1 are evaluated against
a stratified 300-pair gold-standard annotation set (see
Section~\ref{sec:validation}); a mapping is counted as correct only if the
annotator labelled it \emph{correct} (strict; \emph{partial} counts as
false positive). F1 is computed from unrounded P and R; rounding
the displayed values may yield slightly different results.}
\label{tab:baselines}
\small
\begin{tabular}{@{}l r r r r r@{}}
\toprule
\textbf{Method} & \textbf{Mappings} & \textbf{Topics} & \textbf{P} & \textbf{R} & \textbf{F1} \\
\midrule
Jaro--Winkler ($\geq 0.90$) & 937    & 675   & 0.80 & 0.52 & 0.63 \\
TF-IDF cosine               & 4,861  & 2,620 & 0.61 & 0.84 & 0.71 \\
BM25                        & 14,929 & 4,513 & 0.45 & 0.92 & 0.61 \\
BGE-large (ours)            & 16,150 & 4,509 & 0.38 & 1.00 & 0.55 \\
BGE-large ($\geq 0.85$)    & 2,527  & 1,647 & 0.67 & 0.89 & 0.77 \\
\bottomrule
\end{tabular}
\end{table}

The embedding approach captures semantic similarity that lexical methods
cannot: for example, the OpenAlex topic ``Artificial Intelligence in
Medicine'' maps to EDAM's ``Medical informatics'' (cosine similarity 0.87)
and NCIT's ``Biomedical Informatics'' (0.85), neither of which would be
found by any of the three string-based baselines.

\begin{figure}[t]
\centering
\includegraphics[width=\textwidth]{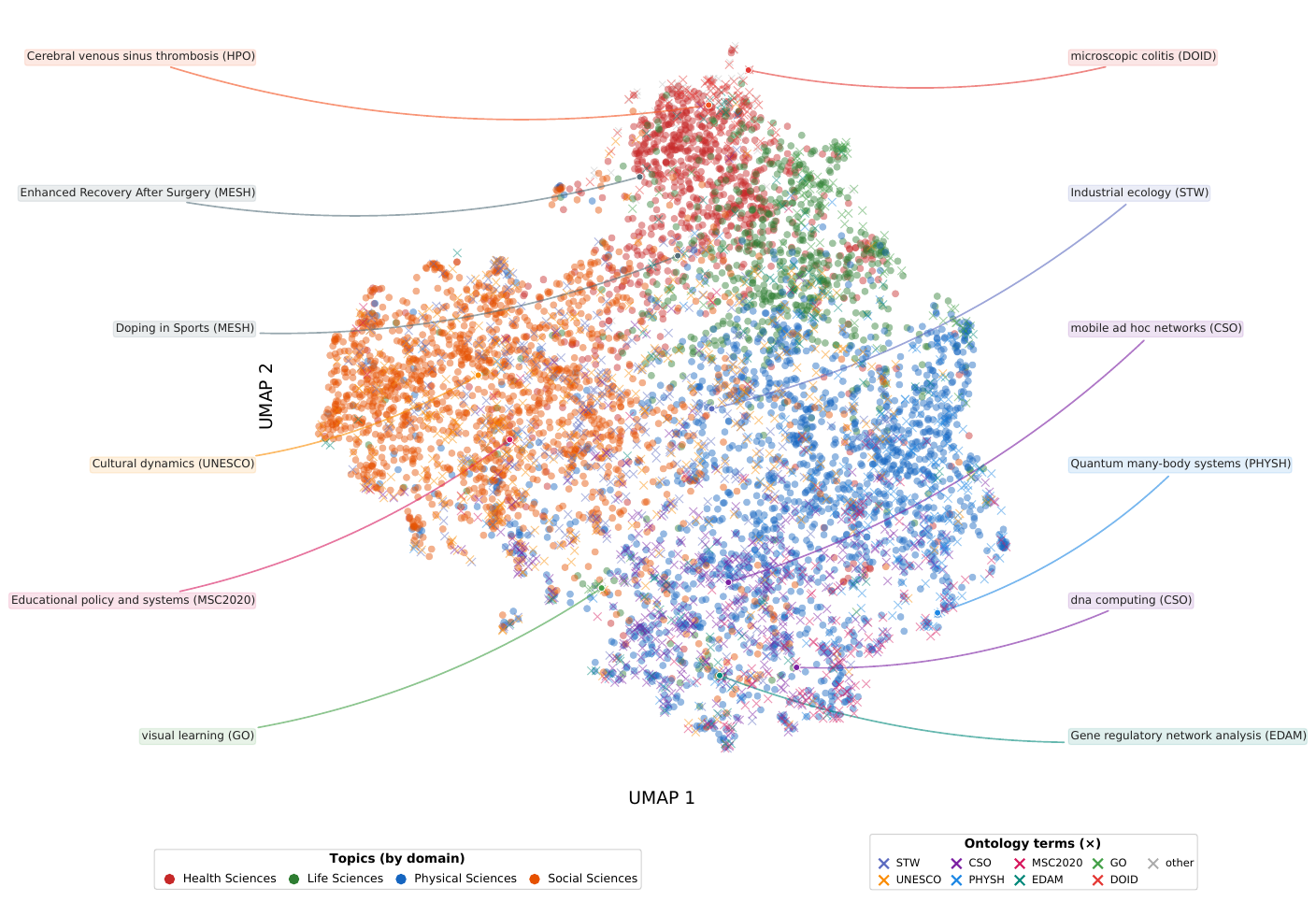}
\caption{UMAP projection of BGE-large embeddings for OpenAlex topics (points)
and matched ontology terms (crosses), colored by OpenAlex domain. Semantic
clusters emerge naturally, with domain-specific ontology terms co-locating with
their corresponding topics.}
\label{fig:umap}
\end{figure}

\begin{figure}[t]
\centering
\includegraphics[width=\textwidth]{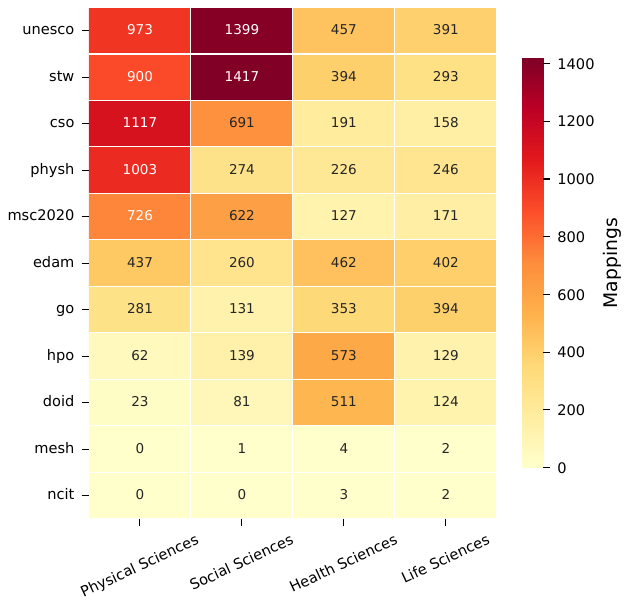}
\caption{Ontology reach heatmap showing the number of high-quality mappings
($\text{similarity} \geq 0.85$) between each ontology and each OpenAlex domain.
The multi-ontology design ensures coverage across all scientific areas.}
\label{fig:heatmap}
\end{figure}

Figure~\ref{fig:umap} visualizes the joint embedding space using
UMAP~\cite{mcinnes2018umap}, showing how OpenAlex topics cluster by domain and
align with terms from domain-specific ontologies. Figure~\ref{fig:heatmap}
displays the ontology-by-domain reach heatmap, confirming that different
ontologies specialize in different scientific areas: MeSH dominates health
sciences, CSO covers computer science, GO spans molecular biology, and AGROVOC
bridges agricultural and environmental sciences.

\section{Data Records}
\label{sec:records}

The Science Data Lake is hosted on HuggingFace Datasets
(\url{https://huggingface.co/datasets/J0nasW/science-datalake},
DOI: \texttt{10.57967/hf/7850}), which provides a persistent
DataCite DOI for citation. Remote users can query the Parquet files directly
through DuckDB's \texttt{hf://} protocol without downloading the full dataset.
Due to licensing restrictions, the HuggingFace-hosted version does not include
Semantic Scholar (distributed under ODC-BY with additional terms of service
restricting use to non-commercial research) or Reliance on Science
(CC~BY-NC~4.0). Both sources can be obtained locally using the provided download
scripts and a free Semantic Scholar API key; once downloaded, the pipeline
automatically integrates them into the unified schema.

The full dataset comprises approximately 960\,GB of compressed Apache Parquet
files organized into 22~schema directories, each containing one or more Parquet
files corresponding to the tables of that schema. The lightweight DuckDB database
file ({\raise.17ex\hbox{$\scriptstyle\sim$}}270\,KB) defines 153~SQL views that
reference these Parquet files and can be regenerated from source using the
provided pipeline scripts.

\begin{minipage}{\textwidth}
The principal schemas and their contents are:

\begin{itemize}[nosep]
  \item \texttt{s2ag}: papers (231M), abstracts, citations, authors, publication
    venues from Semantic Scholar.
  \item \texttt{openalex}: works (479M), authors, institutions, sources, topics,
    concepts, publishers, funders from OpenAlex.
  \item \texttt{sciscinet}: paper metrics (250M), disruption indices, atypicality
    scores, team size indicators from SciSciNet.
  \item \texttt{pwc}: papers (513K), code links, tasks, datasets, methods from
    Papers with Code.
  \item \texttt{retwatch}: retracted papers (69K) with retraction reasons and
    dates from Retraction Watch.
  \item \texttt{ros}: patent--paper citation pairs (47.8M) from Reliance on Science.
  \item \texttt{p2p}: preprint-to-published DOI mappings (146K) from bioRxiv/medRxiv.
  \item \texttt{xref}: cross-source linkage tables---\texttt{unified\_papers}
    (293M, 29~columns), \texttt{doi\_map}, \texttt{topic\_ontology\_map}
    (16,150 mappings), and temporal guardrail views
    (\texttt{source\_temporal\_coverage}, \texttt{paper\_temporal\_flags}).
  \item 13~ontology schemas (e.g., \texttt{mesh}, \texttt{go}, \texttt{cso}),
    each containing \texttt{*\_terms}, \texttt{*\_hierarchy}, and optionally
    \texttt{*\_xrefs} tables.
\end{itemize}

\smallskip
Each source retains its original license: CC0 (OpenAlex), ODC-BY (S2AG),
CC~BY 4.0 (SciSciNet), CC~BY-SA 4.0 (Papers with Code), CC~BY-NC 4.0
(Reliance on Science), and open/public-domain for the remaining sources.
Users should comply with the most restrictive license applicable to the
sources they query.
\end{minipage}

\section{Technical Validation}
\label{sec:validation}

We validated the Science Data Lake through 10~automated sanity checks
(Table~\ref{tab:sanity}), cross-source citation correlation analysis, and manual
inspection of ontology mappings.

\begin{table}[t]
\centering
\caption{Summary of automated sanity checks. All 10 checks passed without
violations across the full dataset.}
\label{tab:sanity}
\small
\begin{tabularx}{\textwidth}{@{}c l c X@{}}
\toprule
\textbf{\#} & \textbf{Check} & \textbf{Result} & \textbf{Detail} \\
\midrule
1  & DOI format (no prefix, lowercase) & PASS & 0 violations / 293M \\
2  & Coverage flags match data presence & PASS & 0 mismatches (OA, S2AG, SSN) \\
3  & Primary key uniqueness (no duplicate DOIs) & PASS & 293,123,121 unique = total \\
4  & OpenAlex ID format \& joinability & PASS & 0 format violations; 69\% topic join \\
5  & Ontology map: no orphan topic IDs & PASS & 0 orphan topic\_ids \\
6  & RoS to OpenAlex join (10K sample) & PASS & 86\% match rate \\
7  & Citation cross-source correlation & PASS & $r = 0.76$--$0.87$ pairwise \\
8  & Year distribution (NULL/invalid) & PASS & NULL: 0.53\%, invalid: 0.001\% \\
9  & Spot-check known papers & PASS & Wakefield retraction flags correct \\
10 & Vignette count reproducibility & PASS & All 4 counts match exactly \\
\bottomrule
\end{tabularx}
\end{table}

\subsection{DOI and Schema Integrity}

Checks~1--5 verify the structural integrity of the cross-reference layer.
The DOI format check (Check~1) confirms that all 293~million entries in
\texttt{unified\_papers} use the canonical lowercase, prefix-free format
with zero violations. Coverage flags (Check~2) are Boolean columns indicating
whether each paper appears in OpenAlex, S2AG, and SciSciNet; all flags
correctly reflect the presence or absence of data in the corresponding
source tables. Primary key uniqueness (Check~3) confirms that each DOI
appears exactly once. The OpenAlex ID format check (Check~4) validates that
all IDs conform to the expected pattern and that 69\% of papers successfully
join to the \texttt{works\_topics} table (the remainder lack topic
assignments in OpenAlex). Check~5 verifies that every topic ID in the
ontology mapping table exists in the OpenAlex topic taxonomy, with zero
orphans found.

\subsection{Cross-Source Citation Agreement}

To assess the consistency of citation counts across databases, we computed
pairwise Pearson correlations for papers present in all three large sources
(S2AG, OpenAlex, SciSciNet; $n \approx 121$M). The correlations are:
S2AG--OpenAlex $r = 0.76$, S2AG--SciSciNet $r = 0.87$, and
OpenAlex--SciSciNet $r = 0.86$. The mean absolute differences are 4.14
(S2AG--OA), 3.84 (S2AG--SSN), and 2.30 (OA--SSN) citations.

\begin{figure}[t]
\centering
\includegraphics[width=\textwidth]{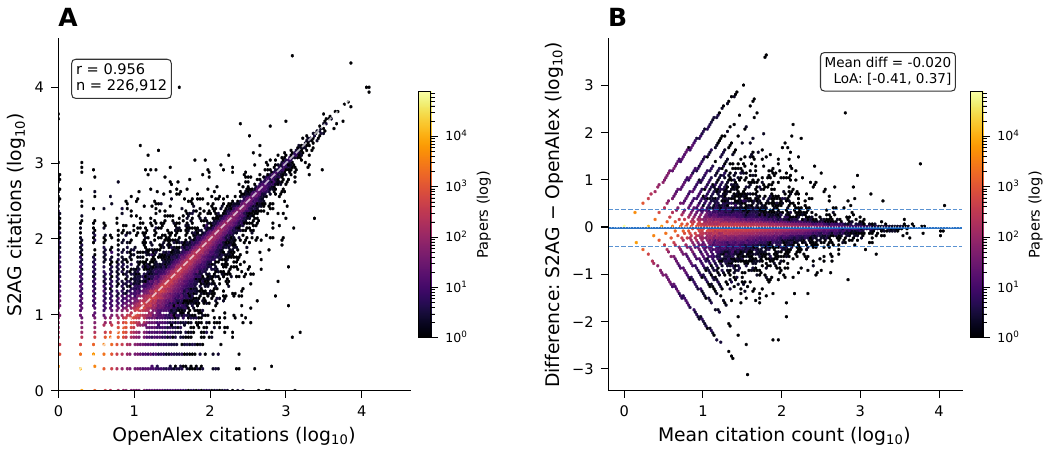}
\caption{Bland--Altman plot of citation count agreement between Semantic Scholar
(S2AG) and OpenAlex. Each point represents a paper; the $x$-axis shows the mean
citation count across both sources, and the $y$-axis shows the difference
(S2AG $-$ OpenAlex). Dashed lines indicate the mean difference and 95\% limits
of agreement.}
\label{fig:blandaltman}
\end{figure}

Figure~\ref{fig:blandaltman} presents the Bland--Altman analysis of citation
agreement between S2AG and OpenAlex. The plot reveals that disagreement
increases with citation magnitude, and identifies systematic outliers---most
notably a single paper with 257,887 citations in S2AG and zero in OpenAlex,
attributable to differences in citation counting methodology and coverage
scope.

The two-of-three correlations exceeding $r = 0.8$ confirm that the three
sources provide broadly consistent citation information, while the
non-negligible disagreements (particularly S2AG--OA at $r = 0.76$) underscore
the value of preserving all three counts for sensitivity analyses.

We deliberately preserve all three independent citation counts rather than
imposing a single resolved value. The six Boolean coverage flags
(\texttt{has\_s2ag}, \texttt{has\_openalex}, \texttt{has\_sciscinet},
\texttt{has\_pwc}, \texttt{has\_retraction}, \texttt{has\_patent}) serve as
implicit conflict-detection indicators: papers present in fewer sources
warrant additional caution when comparing citation-dependent metrics.
Imposing automated conflict resolution would require editorial judgment
about source reliability that may vary by field, era, and research
question---choices we believe should remain with the analyst.

\subsection{Ontology Alignment Validation}

To quantify alignment quality beyond manual inspection, we drew a stratified
sample of 300~mappings from the 16,150 total: 50 from the exact tier
($\geq 0.95$), 100 from high-quality ($0.85$--$0.95$), 100 from mid-range
($0.75$--$0.85$), and 50 from the borderline tier ($0.65$--$0.75$), with
proportional representation of all 13~ontologies within each stratum. Each
pair was independently labelled as \emph{correct} (semantically equivalent),
\emph{partial} (meaningful relationship but not equivalent), or
\emph{incorrect} (unrelated).

\begin{figure}[t]
\centering
\includegraphics[width=0.8\textwidth]{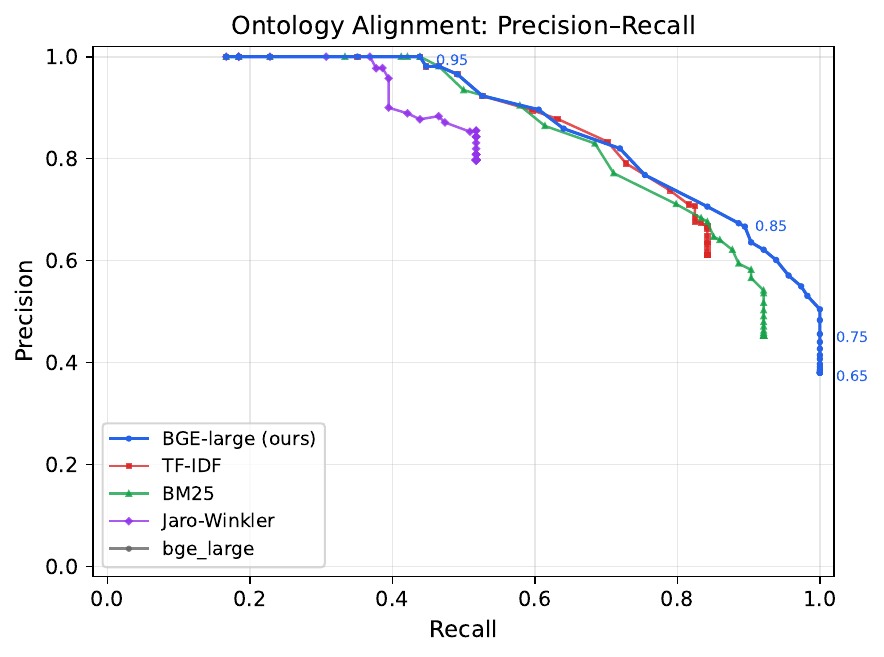}
\caption{Precision--recall curves for the four ontology alignment methods,
swept across cosine-similarity thresholds from 0.60 to 1.00. Annotated
points mark key thresholds.}
\label{fig:precision_recall}
\end{figure}

Under strict evaluation (only \emph{correct} = true positive), per-stratum
precision is: exact tier 1.00 (50/50), high-quality 0.51 (51/100),
mid-range 0.13 (13/100), and borderline 0.00 (0/50). Notably, zero
incorrect mappings were found in the exact and high-quality tiers---all
non-correct labels there are \emph{partial} (meaningful but non-equivalent
relationships). At the recommended $\geq 0.85$ operating point, BGE-large
achieves $P = 0.67$, $R = 0.89$, $F1 = 0.77$. The same gold-standard set
was used to evaluate three lexical baselines (TF-IDF, BM25, Jaro--Winkler;
see Table~\ref{tab:baselines}).
Figure~\ref{fig:precision_recall} shows the precision--recall trade-off
across similarity thresholds for all four methods.

Typical borderline matches in the high-quality tier included
legitimate cross-domain connections such as ``Bioelectronics'' (OpenAlex)
$\to$ ``Biosensors'' (EDAM, similarity 0.856), which reflect meaningful
semantic proximity rather than errors.

\subsection{Temporal Coverage Guardrails}

The temporal coverage of individual sources introduces caveats for
longitudinal analyses. To make these caveats programmatically accessible,
the \texttt{xref} schema includes two auxiliary views.
\texttt{xref.source\_temporal\_coverage} records the year range and
coverage type of each source (e.g., SciSciNet metrics end at approximately
2022, RoS patent citations are strongest through late 2023).
\texttt{xref.paper\_temporal\_flags} provides per-paper Boolean flags:
\texttt{sciscinet\_metrics\_stale} (true when a paper's year exceeds 2022
and the paper has SciSciNet records), \texttt{ros\_coverage\_incomplete}
(true when year exceeds 2023 and the paper has patent citations), and
\texttt{year\_missing} (true when the publication year is NULL). Users can
join these flags to any analysis query to restrict results to temporally
reliable subsets---for instance, filtering out papers with stale disruption
scores before computing field-level disruption trends.

\subsection{Known Limitations}

Users should verify the temporal coverage of each source before drawing
conclusions about recent trends. The OpenAlex topic taxonomy may evolve
across snapshots, potentially affecting ontology mapping stability.

\section{Usage Notes}
\label{sec:usage}

\subsection{Setup}

The Science Data Lake can be deployed by cloning the repository, running the
pipeline via \texttt{datalake\_cli.py} (which downloads, converts, and links
all sources), and connecting to the resulting DuckDB database. For users
without local storage, HuggingFace-hosted Parquet files can be queried
directly through DuckDB's \texttt{httpfs} extension. All queries below use
standard SQL and execute within DuckDB.

\subsection{Vignette 1: Disruption, Code Adoption, and Ontology Landscape}

This vignette examines whether papers that release code exhibit different
disruption profiles than those that do not, and maps this pattern across
ontology-defined domains.

Joining \texttt{sciscinet.paper\_metrics} (disruption index
CD\textsubscript{5}), \texttt{xref.unified\_papers} (code-availability flag
from Papers with Code), and \texttt{xref.topic\_ontology\_map} (ontology
bridging), we identified 139,873 papers with associated code repositories
(0.048\% of the unified table). Papers with code showed a mean
CD\textsubscript{5} of $-0.0005$, compared with $+0.0026$ for papers
without code, suggesting that code-releasing papers tend to be slightly more
consolidating (building on existing work) rather than disruptive. Ontology
mapping reveals that this pattern varies across domains: computer science
topics (mapped via CSO) show the strongest code adoption, while biomedical
topics (mapped via GO and MeSH) show lower code rates but higher disruption
variability.

This analysis is \emph{only possible} because it requires simultaneous access
to disruption scores (SciSciNet), code flags (Papers with Code), topic
assignments (OpenAlex), and ontology bridging (our linkage)---four resources
that exist in no single database.

\subsection{Vignette 2: Retraction Profiles and Ontology Enrichment}

This vignette characterizes retracted papers by their pre-retraction impact
metrics and identifies ontology domains with anomalous retraction rates.

{\emergencystretch=1.5em
Joining \texttt{retwatch.retracted\_papers}, \texttt{sciscinet.paper\_metrics},
and \texttt{xref.unified\_papers}, we obtained 58,775 retracted papers with
associated SciSciNet metrics. Retracted papers show a mean disruption of
$0.0035$ compared with $0.0026$ for non-retracted papers, and the most-cited
retracted paper accumulated 8,062 citations before retraction. Ontology-level
enrichment analysis reveals retraction hotspots: topics mapped to ``AI
Applications'' show 394$\times$ enrichment, and ``Advanced Technology''
topics show 338$\times$ enrichment relative to baseline retraction rates.\par}

This analysis requires retraction flags (Retraction Watch), disruption
scores (SciSciNet), citation counts (OpenAlex), and ontology mapping---a
combination unavailable in any single source.

\subsection{Vignette 3: Patent Impact and Multi-Ontology Footprint}

This vignette quantifies the citation and impact characteristics of papers
cited by patents, broken down by ontology domain.

Joining \texttt{ros.patent\_paper\_pairs}, \texttt{xref.unified\_papers},
and \texttt{xref.topic\_ontology\_map}, we identified 312,929 patent-cited
papers (0.107\% of the unified table). These papers have dramatically
higher impact: mean citation count of 94.3 versus 16.1 for non-patent-cited
papers (5.8$\times$), and mean FWCI of 4.7 versus 1.5 (3.1$\times$). The
multi-ontology footprint reveals that patent-cited papers cluster in applied
domains: MeSH-mapped health science topics, CSO-mapped computer science
topics, and ChEBI-mapped chemistry topics dominate.

The temporal coverage caveat applies: RoS patent citations are strongest
through late 2023 due to patent processing lag, so very recent papers may
have incomplete patent linkage.

\subsection{Vignette 4: Cross-Source Citation Reliability}

This vignette demonstrates record-level citation comparison across three
independent databases.

Restricting to the 121~million papers present in all three large sources
(S2AG, OpenAlex, SciSciNet), we computed pairwise citation correlations:
S2AG--OpenAlex $r = 0.76$, S2AG--SciSciNet $r = 0.87$, and OpenAlex--SciSciNet
$r = 0.86$. The mean absolute differences range from 2.3 to 4.1~citations.
Relative disagreement is most pronounced for low-cited papers (mean relative
difference $\sim$20\% for papers with $<$10 citations) and diminishes for
high-cited papers. The most extreme outlier---a paper with 257,887 citations
in S2AG and zero in OpenAlex---illustrates how coverage and methodology
differences can produce dramatic record-level discrepancies.

Whereas Section~\ref{sec:validation} uses these correlations to validate
data integrity, this vignette illustrates the analytical workflow
itself---showing how researchers can diagnose source-specific biases and
select the citation count most appropriate for their study design. The
three-way comparison requires parallel citation counts from independent
sources to coexist in a single queryable table---a configuration that
would be impractical to assemble through pairwise API-based comparisons
at this scale.

\begin{figure}[t]
\centering
\includegraphics[width=\textwidth]{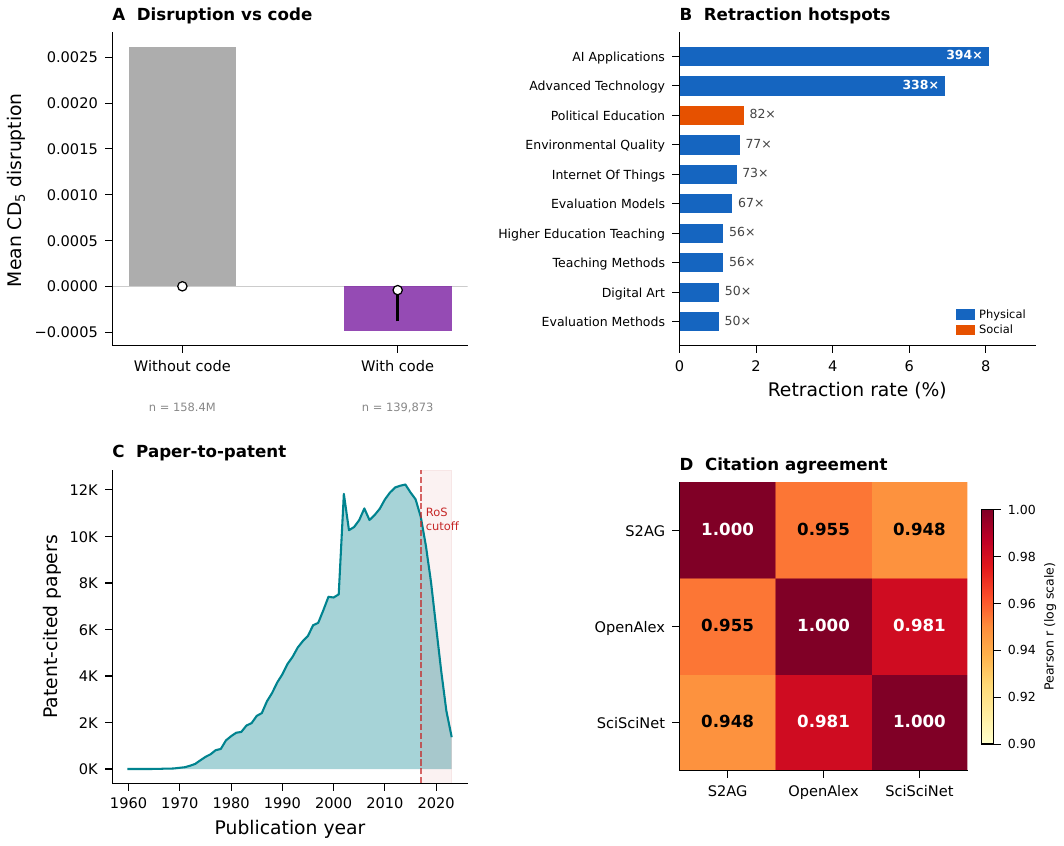}
\caption{Composite vignette results (2$\times$2 panels). Top-left: disruption
distributions for papers with versus without code (Vignette~1). Top-right:
retraction enrichment by ontology domain (Vignette~2). Bottom-left: citation
distributions for patent-cited versus non-patent-cited papers (Vignette~3).
Bottom-right: pairwise citation agreement across three sources (Vignette~4).}
\label{fig:vignettes}
\end{figure}

Figure~\ref{fig:vignettes} summarizes the results of all four vignettes.

\subsection{AI-Assisted Querying}

The Science Data Lake includes a structured schema reference
(\texttt{SCHEMA.md}, approximately 1,200~lines) designed to serve as
context for large language model (LLM) based coding agents. This
document provides every table name, column, data type, row count, and
size tier, along with nine cross-dataset join strategies and common query
recipes---all in a format optimized for LLM consumption rather than
narrative reading. For example, a query such as ``find the most disruptive
papers in computer science that have open-source code and check their
retraction status'' requires joining four schemas (\texttt{sciscinet},
\texttt{xref}, \texttt{pwc}, \texttt{retwatch}) with appropriate DOI
normalization; \texttt{SCHEMA.md} consolidates the metadata needed to
formulate such joins.

Rather than building a custom natural-language interface (which would
require maintenance and constrain the query space), we provide structured
documentation that any general-purpose LLM agent can consume. As LLM
capabilities evolve, the same schema reference remains useful without
modification. Systematic evaluation of text-to-SQL performance across
LLM providers is beyond the scope of this data descriptor.

\subsection{Limitations and Extensibility}

The Science Data Lake inherits the limitations of its constituent sources.
Temporal coverage varies: SciSciNet metrics end around 2022, RoS exhibits
patent processing lag, and OpenAlex snapshot dates may trail real-time data by
weeks to months. Papers without DOIs (estimated at 5--15\% depending on field
and era) are excluded from cross-source linkage. The ontology mapping relies
on the current OpenAlex topic taxonomy, which may evolve across snapshots.

The architecture is designed for extensibility: adding a new data source
requires writing a Parquet converter and registering the schema in the
pipeline configuration. Community contributions of additional sources,
ontologies, or cross-reference methods are encouraged.

\section{Code Availability}
\label{sec:code}

All code for constructing and querying the Science Data Lake is available in a
public GitHub repository (\url{https://github.com/J0nasW/science-datalake}).
The pipeline is implemented in Python~3.12 with the
following principal dependencies: DuckDB~1.4.2~\cite{raasveldt2019duckdb},
PyArrow~22.0, and sentence-transformers~5.2.2 (for ontology alignment).

The pipeline comprises five stages, each implemented as a subcommand of the
master CLI script:

\begin{enumerate}[nosep]
  \item \textbf{Download} (\texttt{datalake\_cli.py download}): retrieves
    source snapshots from their official distribution points (S3 buckets,
    APIs, direct downloads).
  \item \textbf{Convert} (\texttt{datalake\_cli.py convert}): transforms
    each source from its native format (JSON Lines, CSV, N-Triples) into
    columnar Apache Parquet files.
  \item \textbf{Create views} (\texttt{create\_unified\_db.py}): generates
    the DuckDB database with 153~SQL views across 22~schemas.
  \item \textbf{Materialize} (\texttt{materialize\_unified\_papers.py}):
    constructs the \texttt{xref.unified\_papers} join table through DOI
    normalization and multi-source record linkage.
  \item \textbf{Build linkage} (\texttt{build\_embedding\_linkage.py}):
    computes BGE-large embeddings for ontology terms and OpenAlex topics,
    builds a FAISS index, and produces the ontology alignment table
    \texttt{xref.topic\_ontology\_map}.
\end{enumerate}

Additional scripts include \texttt{convert\_ontologies.py} (five
format-specific parsers for the 13~ontologies) and
\texttt{ontology\_registry.py} (a declarative registry of ontology URLs
and formats). The repository also includes a structured schema reference
(\texttt{SCHEMA.md}) that documents all 153~views with their columns,
types, row counts, and cross-dataset join strategies, designed to be
consumed by both LLM-based coding agents and human developers.

\section*{Acknowledgements}
The author thanks Hamburg University of Technology (TUHH) for institutional
support. Computational resources were provided by the TUHH High-Performance
Computing cluster. The author gratefully acknowledges the OpenAlex team for
providing open, freely accessible data snapshots via Amazon Web Services S3,
and the teams behind Semantic Scholar, SciSciNet, Papers with Code,
Retraction Watch, Reliance on Science, and the PreprintToPaper project for
making their datasets openly available.

\section*{Author Contributions}
\textbf{J.W.}: Conceptualization, Methodology, Software, Data curation,
Validation, Formal analysis, Investigation, Visualization,
Writing~---~original draft, Writing~---~review \& editing.

\section*{Competing Interests}
The author declares no competing interests.


\end{document}